**RESEARCH** **Open Access**

# Adaptive weight-modified Riesz mean filter for high-density salt-and-pepper noise removal

Md Jahidul Islam[1*]
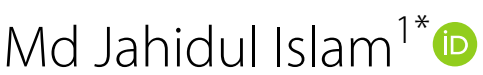

*Correspondence:
2006123@eee.buet.ac.bd

[1] Department of Electrical and Electronics Engineering, Bangladesh University of Engineering and Technology, Dhaka, Bangladesh

## Abstract

This paper introduces a novel filter, the adaptive weight-modified Riesz mean filter (AWMRmF), designed for the effective removal of high-density salt-and-pepper noise. AWMRmF integrates a pixel weight function and adaptivity condition inspired by the different adaptive modified Riesz mean filter. In this study, the performance of AWMRmF was evaluated against established filters such as adaptive frequency median filter, adaptive weighted mean filter, adaptive Ces´aro mean filter, adaptive Riesz mean filter, and improved adaptive weighted mean filter. The assessment was conducted on 26 typical test images, varying noise levels from 60 to 95%. The findings indicate that, in terms of both peak signal-to-noise ratio and structural similarity metrics, AWMRmF outperformed other state-of-the-art filters. Furthermore, AWMRmF demonstrated superior performance in mean PSNR and SSIM results as well.

**Keywords:** Salt-and-pepper noise, Non-linear functions, Noise removal, Matrix algebra, Image denoising, Riesz mean

## Introduction

In the era of advancing technology, the acquisition of diverse image types, ranging from medical images [1] to astronomical images [2], and satellite images [3], has become increasingly prevalent. However, the transmission and acquisition processes of these images are susceptible to a phenomenon known as noise, which compromises image quality. Noise manifests in various forms, such as Gaussian noise, impulse noise, and speckle noise, each impacting image quality differently.

This work focuses on salt-and-pepper noise (SPN), a form of fixed-valued impulse noise. SPN, characterized by white (255) and black (0) dots in images, stems from factors such as sensor issues, electrical conditions, and transmission errors. To mitigate the detrimental effects of SPN, a range of image denoising filters have been proposed, each designed to address specific challenges associated with SPN.

Pioneer approaches such as the standard median filter (SMF) [4] and adaptive median Filter (AMF) [5] have laid the foundation for SPN removal. However, recent advancements have introduced innovative filters like the adaptive weighted mean filter (AWMF) [6] and improved adaptive weighted mean filter (IAWMF) [7], different adaptive modified Riesz mean filter (DAMRmF) [8], each aiming to enhance the denoising process.

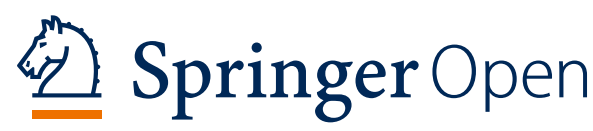

This paper introduces a novel high-density SPN filter, named adaptive weight-modified Riesz mean filter (AWMRmF). Building upon the adaptivity concept of ARmF and the pixel similarity principle of DARMmF, AMRmF surpasses existing filters in terms of peak signal-to-noise ratio (PSNR) and structural similarity (SSIM) [9]. Through a comprehensive experimental study, AWMRmF demonstrates superior performance compared to established filters like AFMF [10], AWMF, ACmF [11], ARmF [12], and IAWMF across various noise levels.

The subsequent sections of this paper delve into the fundamental definitions and notations, present the proposed AwMRmF filter in detail, conduct an experimental study to validate its effectiveness, and conclude with discussions and remarks for further research. The innovative combination of adaptivity and pixel similarity in AWMRmF promises to contribute significantly to the field of image denoising, addressing the challenges posed by high-density salt-and-pepper noise in diverse image domains.

## Basic definitions and notations

**Definition 1** Let A:= $[a_{ij}]_{m\times n}$ be an image matrix (IM) such that aij is an unsigned integer number and $0 \leq a_{ij} \leq 255$. Then, $a_{ij}$ is called a noisy entry of A if $a_{ij} = 0$ or $a_{ij} = 255$; otherwise, $a_{ij}$ is called a regular entry of A.

**Definition 2** Let A be an IM. Then, A is called a noise image matrix (NIM) if for some i and j, $a_{ij}$ is a noisy entry of A.

**Definition 3** Let A:= $[a_{ij}]_{m\times n}$ and t∈{1, 2, ..., min(m, n)}. Then, the matrix $[a_{ij}]_{(m+2t)\times(n+2t)}$ called t-symmetric pad matrix of A is denoted by $\overline{\overline{\mathrm{A}}}_{\text{t-sym}}$ (or briefly$\overline{\overline{A_t}}$) and is defined as follows:

$$\begin{bmatrix} a_{tt} & \cdots & a_{t1} & a_{t1} & a_{t2} & \cdots & a_{tn} & a_{tn} & \cdots & a_{t(n-t+1)} \\ \vdots & \ddots & \vdots & \vdots & \vdots & \ddots & \vdots & \vdots & \ddots & \vdots \\ a_{1t} & \cdots & a_{11} & a_{11} & a_{12} & \cdots & a_{1n} & a_{1n} & \cdots & a_{1(n-t+1)} \\ a_{1t} & \cdots & a_{11} & \boldsymbol{a_{11}} & \boldsymbol{a_{12}} & \cdots & \boldsymbol{a_{1n}} & a_{1n} & \cdots & a_{1(n-t+1)} \\ a_{2t} & \cdots & a_{21} & \boldsymbol{a_{21}} & \boldsymbol{a_{22}} & \cdots & \boldsymbol{a_{2n}} & a_{2n} & \cdots & a_{2(n-t+1)} \\ a_{3t} & \cdots & a_{31} & \boldsymbol{a_{31}} & \boldsymbol{a_{32}} & \cdots & \boldsymbol{a_{3n}} & a_{3n} & \cdots & a_{3(n-t+1)} \\ \vdots & \ddots & \vdots & \vdots & \vdots & \ddots & \vdots & \vdots & \ddots & \vdots \\ a_{mt} & \cdots & a_{m1} & \boldsymbol{a_{m1}} & \boldsymbol{a_{m2}} & \cdots & \boldsymbol{a_{mn}} & a_{mn} & \cdots & a_{m(n-t+1)} \\ a_{mt} & \cdots & a_{m1} & a_{m1} & a_{m2} & \cdots & a_{mn} & a_{mn} & \cdots & a_{m(n-t+1)} \\ \vdots & \ddots & \vdots & \vdots & \vdots & \ddots & \vdots & \vdots & \ddots & \vdots \\ a_{(m-t+1)t} & \cdots & a_{(m-t+1)1} & a_{(m-t+1)1} & a_{(m-t+1)2} & \cdots & a_{(m-t+1)n} & a_{(m-t+1)n} & \cdots & a_{(m-t+1)(n-t+1)} \end{bmatrix} \tag{1}$$

**Definition 4** Let A be an IM. Then, the matrix B:= $[b_{ij}]_{m\times n}$ is called a binary matrix of A where

$$b_{ij} = \begin{cases} 0, & \text{if } a_{ij} \text{ is a noisy entry of } A \\ 1, & \text{otherwise} \end{cases} \tag{2}$$

***Example 1***

*Let,* $A := \begin{bmatrix} 0 & 85 & 76 \\ 35 & 255 & 255 \\ 0 & 150 & 73 \end{bmatrix}$ *Then,* $\overline{\overline{A_2}} := \begin{bmatrix} 255 & 35 & 35 & 255 & 255 & 255 & 255 \\ 85 & 0 & 0 & 85 & 76 & 76 & 85 \\ 85 & 0 & 0 & 85 & 76 & 76 & 85 \\ 255 & 35 & 35 & 255 & 255 & 255 & 255 \\ 150 & 0 & 0 & 150 & 73 & 73 & 150 \\ 150 & 0 & 0 & 150 & 73 & 73 & 150 \\ 255 & 35 & 35 & 255 & 255 & 255 & 255 \end{bmatrix}_{7\times 7}$

**Definition 5** Let A := $[a_{ij}]_{m\times n}$ and k ∈ {1, 2, ..., t}. Then, the matrix.

$$\begin{bmatrix} \overline{\overline{a_{(i+t-k)(j+t-k)}}} & \cdots & \overline{\overline{a_{(i+t-k)(j+t+k)}}} \\ \vdots & \overline{\overline{a_{(i+t)(j+t)}}} & \vdots \\ \overline{\overline{a_{(i+t+k)(j+t-k)}}} & \cdots & \overline{\overline{a_{(i+t+k)(j+t+k)}}} \end{bmatrix} \tag{3}$$

is called k-approximate matrix of $a_{ij}$ in At and is denoted by $A_{ij}^{k}$.

***Example 2***

*Let's consider Example 1. Then,* $A_{43}^{2} = \begin{bmatrix} \overline{\overline{a_{55}}} & \overline{\overline{a_{56}}} & \overline{\overline{a_{57}}} \\ \overline{\overline{a_{65}}} & \overline{\overline{a_{66}}} & \overline{\overline{a_{67}}} \\ \overline{\overline{a_{75}}} & \overline{\overline{a_{76}}} & \overline{\overline{a_{77}}} \end{bmatrix} = \begin{bmatrix} 73 & 73 & 150 \\ 73 & 73 & 150 \\ 255 & 255 & 255 \end{bmatrix}$

**Definition 6** A matrix with all zero entries is called a zero or null matrix and denoted by [0].

**Definition 7** Let A be an IM. Then, the value $ps\left(a_{ij,}a_{st}\right) := \left(\frac{1}{1+|i-s|+|j-t|}\right)^{2}$ is called pixel similarity between $a_{ij}$ and $a_{st}$.

**Definition 8** Let A be an NIM. Then the value

$$Rm(A_{ij}^{k}) = \frac{\sum_{(s,t)\in I_{ij}^{k}} ps(a_{st}, a_{(k+1)(k+1)})a_{st}}{\sum_{(s,t)\in I_{ij}^{k}} ps(a_{st}, a_{(k+1)(k+1)})} \tag{4}$$

is called Riesz mean of$A_{ij}^{k}$. Here $I_{ij}^{k} := \left\{(s,t) : a_{st} \text{ is a regular entry of } A_{ij}^{k}\right\}$.

**Definition 9** Let A be an NIM. Then, the value $pw(a_{st}, k) := \left(\frac{1}{1+(k+1-s)^{2}+(k+1-t)^{2}}\right)^{2}$ is called pixel weight of $a_{st}$ in $A_{ij}$ and the set of all indexes of the regular pixels in$A_{ij}^{k}$, respectively.

**Definition 10** Let A be an NIM. Then the value

$$MRm(A_{ij}^{k}) = \frac{\sum_{(s,t)\in I_{ij}^{k}} pw(a_{st}, a_{k})a_{st}}{\sum_{(s,t)\in I_{ij}^{k}} ps(a_{st}, a_{k})} \tag{5}$$

is called modified Riesz mean of$A_{ij}^{k}$. Here $I_{ij}^{k} := \{(s,t) : a_{st} isaregularentryofA_{ij}^{k}\}$.

**Definition 11** Let A be an NIM. Then, the value.

$$WMRm(A_{ij}^k) = \frac{\sum_{(s,t)\in I_{ij}^k} pw(a_{st}, a_k)a_{st}}{\sum_{(s,t)\in I_{ij}^k} ps(a_{st}, a_k)} \quad (6)$$

is called weight-modified Riesz mean function.

$pw(a_{st}, k) := \left(\frac{1}{1+4^{k+1}(k+1-s)^2+4^{k+1}(k+1-t)^2}\right)^2$ is called modified pixel weight of $a_{st}$ in $A_{ij}$ and the set of all indexes of the regular pixels in $A_{ij}^k$, respectively.

The pixel weight-modified function related with the adaptivity condition simultaneously is defined to deal with the high-density SPN. It produces different weights at each time and these weights are more efficient than those of pixel similarity in ARmF and DAMRmF. Thus, AWMRmF using the modified pixel weight performed better than the state-of-the-art filters in high-density SPN for 26 typical images. The test dataset consists of 26 images selected from the public online repository Pixabay (pixabay.com). Instead of using a standard benchmark dataset, these images were chosen to evaluate the algorithm's performance on a diverse set of contemporary, high-resolution, real-world color photographs. The collection is varied in subject matter, including portraits, animals, plants, and complex scenes, and features a wide range of textures, lighting conditions, and detail levels.

## Proposed salt-and-pepper noise denoising method

### Algorithm

The pseudocodes of ARmF, DAMRmf and the proposed filter (AWMRmF) are:

**Algorithm 1:** Adaptive Riesz mean filter (ARmF)

**Input**: NIM A := $[a_{ij}]_{m\times n}$ such that min{m, n} ≥ 5
**Output:** Denoised A := $[a_{ij}]_{m\times n}$
**1.** Convert A from uint8 form to double form
**2.** **for** t = 5 to 1 **do**
**3.** Compute the binary matrix B := $[b_{ij}]_{m\times n}$ of A
**4.** Compute $\overline{\overline{A_t}}$ and $\overline{\overline{B_t}}$
**5.** **for all** i and j **do**
**6.** **if** $b_{ij}$= 0 **then**
**7.** **for** k = 1 **to t do**
**8.** **if** $B_{ij}^k \neq [0]$ **then**
**9.** $a_{ij} \leftarrow \text{Rm}(A_{ij}^k)$
**10.** break
**11.** **end if**
**12.** **end for**
**13.** **end if**
**14.** **end for**
**15.** **end for**
**16.** Convert final denoised image from double back to uint8
**17.** **return** A

**Algorithm 2**: Different adaptive modified Riesz mean filter (DAMRmF)

**Input**: NIM A := $[a_{ij}]_{m\times n}$ such that min{m, n} ≥ 5
**Output:** Denoised A := $[a_{ij}]_{m\times n}$

1. Convert A from uint8 form to double form
2. **for** t = 5 to 1 **do**
3. Compute the binary matrix B := $[b_{ij}]_{m\times n}$ of A
4. Compute $\overline{\overline{A_t}}$ and $\overline{\overline{B_t}}$
5. **for all** i and j **do**
6. **if** $b_{ij}$= 0 **then**
7. **for** k = 1 **to t do**
8. **if** (0 < med($A_{ij}^k$) < 255) AND ($a_{ij}$= 0) OR ($a_{ij}$ = 255) **then**
9. $a_{ij}$ ← MRm($A_{ij}^k$)
10. break
11. **end if**
12. **end for**
13. **end if**
14. **end for**
15. **end for**
16. Convert final denoised image from double back to uint8
17. **return** A

The proposed algorithm of AWMRmF is described as follows. Here, I defined weight-modified Riesz mean (WMRm). AWMRmF is designed to perform the removal of high-density SPN. The flowchart of the algorithm is in Fig. 1 which helps to visualize the algorithm.

**Algorithm 3:** Adaptive weight-modified Riesz mean filter (AWMRmF)

**Input**: NIM A := $[a_{ij}]_{m\times n}$ such that min{m, n} ≥ 5
**Output:** Denoised A := $[a_{ij}]_{m\times n}$

1. Convert A from uint8 form to double form
2. **for** t = 6 to 1 **do**
3. Compute the binary matrix B := $[b_{ij}]_{m\times n}$ of A
4. Compute $\overline{\overline{A_t}}$ and $\overline{\overline{B_t}}$
5. **for all** i and j **do**
6. **if** $b_{ij}$= 0 **then**
7. **for** k = 1 **to t do**
8. **if** ($a_{ij}$= 0) OR ($a_{ij}$ = 255) **then**
9. $a_{ij}$ ← WMRm($A_{ij}^k$)
10. break
11. **end if**
12. **end for**
13. **end if**
14. **end for**
15. **end for**
16. Convert final denoised image from double back to uint8
17. **return** A

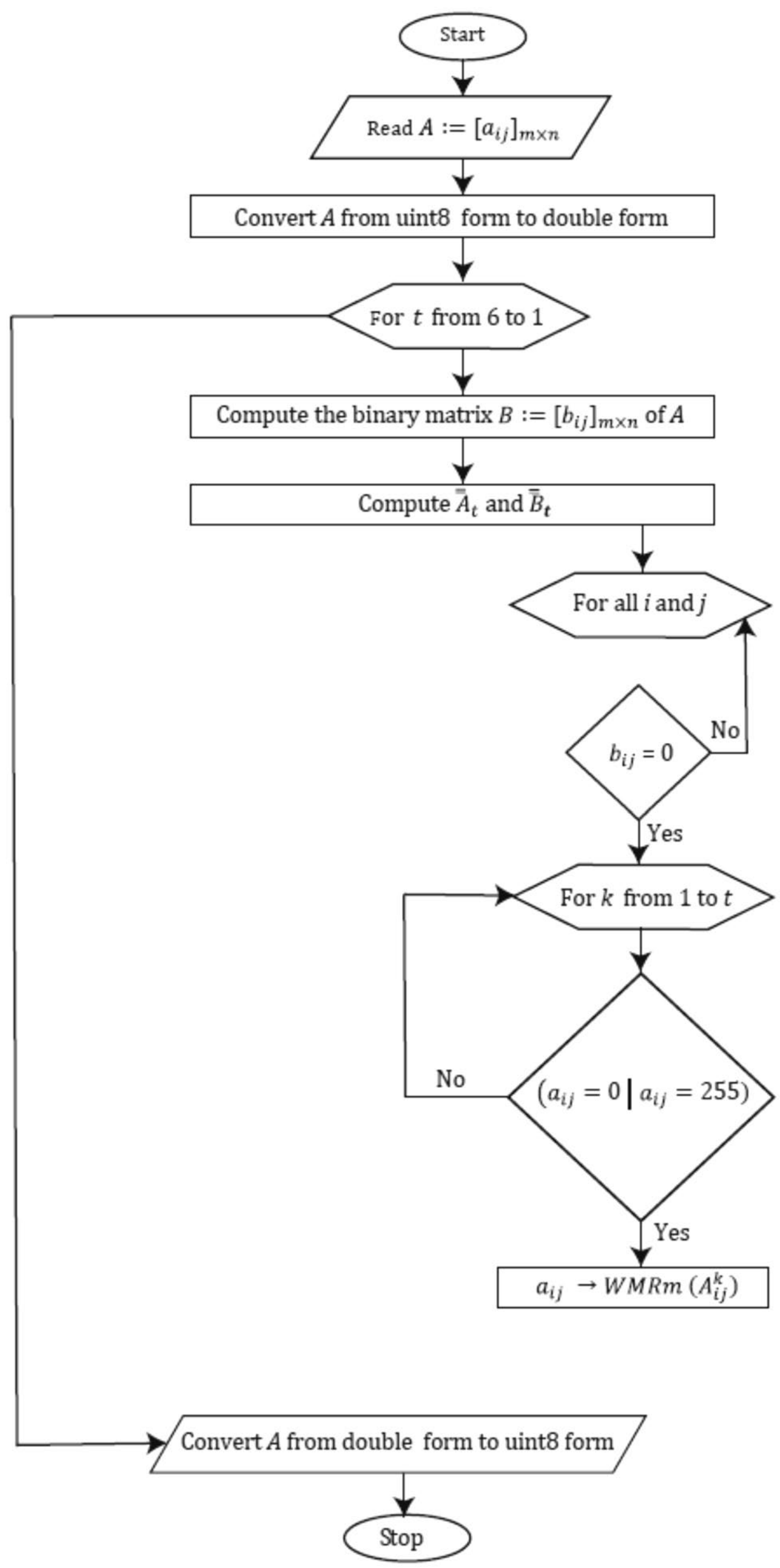

**Fig. 1** Flowchart of the proposed AWMRmF algorithm

## Experimental study

### Image quality assessment matrices

To evaluate image quality, I used peak signal-to-noise ratio (PSNR) and structural similarity (SSIM): Let X:=$[x_{ij}]$ and

Y:=$[y_{ij}]$ be the original image and denoised image, respectively.

$$\mathrm{PSNR}(X, Y) := 10\log_{10}\left(\frac{255^2}{\mathrm{MSE}(X, Y)}\right) \tag{7}$$

**Table 1** Mean PSNR results for 26 typical images with SPN ratios varying from 60 to 95%

| Methods | 60% | 65% | 70% | 75% | 80% | 85% | 90% | 95% | Mean |
|---|---|---|---|---|---|---|---|---|---|
| AFMF | 25.448 | 24.898 | 24.268 | 23.688 | 23.065 | 22.072 | 19.283 | 13.667 | 22.04867 |
| AWMF | 27.500 | 26.866 | 26.167 | 25.438 | 24.673 | 23.775 | 22.722 | 21.155 | 24.78696 |
| ACmF | 27.507 | 26.847 | 26.131 | 25.382 | 24.625 | 23.718 | 22.675 | 21.107 | 24.74904 |
| ARmF | 27.714 | 27.018 | 26.266 | 25.482 | 24.693 | 23.757 | 22.692 | 21.111 | 24.84175 |
| IAWMF | 26.944 | 26.323 | 25.648 | 24.945 | 24.210 | 23.351 | 22.374 | 20.933 | 24.34094 |
| DAMRmF | 27.248 | 26.692 | 26.099 | 25.519 | 24.826 | 24.072 | 23.122 | 21.499 | 24.88471 |
| AWMRmF | 28.040 | 27.436 | 26.787 | 26.094 | 25.373 | 24.536 | 23.554 | 21.981 | 25.47496 |

where MSE(X, Y) represents the Mean Square Error, and it is defined by

$$\mathrm{MSE}(X,Y) := \frac{1}{mn}\sum_{i=1}^{m}\sum_{j=1}^{n}\left(x_{ij}-y_{ij}\right)^2 \tag{8}$$

$$\mathrm{SSIM}(X,Y) := \frac{\left(2\mu_x\mu_y + C_1\right)\left(2\sigma_{xy} + C_2\right)}{\left(\mu_x^2 + \mu_y^2 + C_1\right)\left(\sigma_x^2 + \sigma_y^2 + C_2\right)} \tag{9}$$

where $\mu_x, \mu_y, \sigma_x, \sigma_y$ and $\sigma_{xy}$ are the average intensities, standard deviations, and cross-covariance of images X and Y, respectively.

Additionally, $C_1 = K_1L^2$ and $C_2 = K_2L^2$ are two constants such that $K_1 = 0.01$, $K_2 = 0.03$ and $L = 255$ for 8-bit grayscale images.

### Simulation results

The experiments of the proposed method for SPN denoising were conducted on MATLAB R2023b. Simulations were conducted on a system with an Intel(R) Core(TM) i7-8665U CPU @ 1.90 GHz 2.11 GHz, 16 GB RAM (15.8 GB usable), and Windows 11 Pro.

Table 1 represents the mean PSNR results for 26 typical images affected by high-density SPN. Among the filters considered, AWMRmF demonstrated better performance, achieving higher PSNR values.

Table 2 represents the mean SSIM results for the same set of images. Once again, the AWMRmF filter outperformed the others, indicating its effectiveness in preserving the structural integrity of the images amidst high-density SPN.

Table 3 provides a detailed comparison of the PSNR results for some typical images with high-density SPN.

Table 4 represents the SSIM results of the filters for some typical images with high-density SPN.

Table 5 represents the mean execution time (seconds) of the mentioned methods during the simulations. Even though AWMF, ACmF, and ARmF ran in less time than AWMRmF, AWMRmF performed better in denoising the SPN from the pictures. Moreover, at first, DAMRmF performed faster than AWMRmF, but with the increase of the SPN density DAMRmF slowed down and AWMRmF got faster than AWMRmF. Thus the mean execution time of AWMRmF was lower than DAMRmF.

**Table 2** Mean SSIM results for 26 typical images with SPN ratios varying from 60 to 95%

| Methods | 60% | 65% | 70% | 75% | 80% | 85% | 90% | 95% | Mean |
|---|---|---|---|---|---|---|---|---|---|
| AFMF | 0.8427 | 0.8219 | 0.7974 | 0.7697 | 0.7365 | 0.6917 | 0.5986 | 0.3397 | 0.6998 |
| AWMF | 0.8409 | 0.8235 | 0.8032 | 0.7797 | 0.7515 | 0.7158 | 0.6672 | 0.5922 | 0.7467 |
| ACmF | 0.8427 | 0.8248 | 0.8041 | 0.7803 | 0.7518 | 0.7160 | 0.6669 | 0.5921 | 0.7474 |
| ARmF | 0.8487 | 0.8306 | 0.8094 | 0.7849 | 0.7554 | 0.7177 | 0.6668 | 0.5851 | 0.7498 |
| IAWMF | 0.8859 | 0.8684 | 0.8482 | 0.8241 | 0.7950 | 0.7574 | 0.7083 | 0.6349 | 0.7903 |
| DAMRmF | 0.8999 | 0.8847 | 0.8679 | 0.8476 | 0.8224 | 0.7898 | 0.7441 | 0.6655 | 0.8152 |
| AWMRmF | 0.9020 | 0.8872 | 0.8701 | 0.8498 | 0.8253 | 0.7938 | 0.7505 | 0.6800 | 0.8198 |

**Table 3** Mean PSNR results for some typical images with SPN ratios varying from 60 to 95%

| Images | Methods | 60% | 65% | 70% | 75% | 80% | 85% | 90% | 95% | Mean |
|---|---|---|---|---|---|---|---|---|---|---|
| | AFMF | 30.840 | 29.940 | 29.168 | 28.504 | 27.472 | 25.909 | 21.910 | 14.672 | 26.052 |
| | AWMF | 32.164 | 31.354 | 30.595 | 29.706 | 28.767 | 27.606 | 26.259 | 24.164 | 28.827 |
| | ACmF | 32.290 | 31.415 | 30.636 | 29.731 | 28.778 | 27.613 | 26.258 | 24.164 | 28.861 |
| Lena | ARmF | 32.511 | 31.598 | 30.776 | 29.850 | 28.852 | 27.667 | 26.285 | 24.171 | 28.964 |
| | IAWMF | 32.684 | 31.835 | 31.061 | 30.214 | 29.267 | 28.139 | 26.818 | 24.816 | 29.354 |
| | DAMRmF | 32.715 | 31.870 | 31.100 | 30.306 | 29.346 | 28.208 | 26.943 | 24.684 | 29.396 |
| | AWMRmF | 32.616 | 31.839 | 31.130 | 30.331 | 29.412 | 28.346 | 27.135 | 25.070 | 29.485 |
| | AFMF | 28.818 | 28.237 | 27.455 | 26.895 | 26.322 | 25.450 | 21.457 | 14.963 | 24.950 |
| | AWMF | 29.762 | 29.327 | 28.442 | 27.818 | 27.101 | 26.237 | 25.278 | 23.824 | 27.223 |
| | ACmF | 29.837 | 29.372 | 28.482 | 27.828 | 27.106 | 26.236 | 25.280 | 23.825 | 27.246 |
| Bird in the lake | ARmF | 30.118 | 29.599 | 28.650 | 27.955 | 27.199 | 26.301 | 25.315 | 23.827 | 27.370 |
| | IAWMF | 30.334 | 29.817 | 28.944 | 28.286 | 27.561 | 26.713 | 25.783 | 24.390 | 27.728 |
| | DAMRmF | 30.316 | 29.813 | 28.978 | 28.348 | 27.684 | 26.856 | 25.931 | 24.202 | 27.766 |
| | AWMRmF | 30.436 | 29.977 | 29.167 | 28.546 | 27.878 | 27.061 | 26.140 | 24.516 | 27.965 |
| | AFMF | 22.818 | 22.245 | 21.777 | 21.234 | 20.748 | 19.968 | 18.404 | 13.981 | 20.147 |
| | AWMF | 23.553 | 22.962 | 22.430 | 21.778 | 21.123 | 20.388 | 19.624 | 18.520 | 21.297 |
| | IAWMF | 23.907 | 23.304 | 22.801 | 22.135 | 21.493 | 20.796 | 20.064 | 19.005 | 21.688 |
| House | ARmF | 23.764 | 23.114 | 22.561 | 21.860 | 21.166 | 20.416 | 19.632 | 18.519 | 21.379 |
| | ACmF | 23.593 | 22.982 | 22.451 | 21.784 | 21.126 | 20.387 | 19.623 | 18.519 | 21.308 |
| | DAMRmF | 23.958 | 23.379 | 22.896 | 22.272 | 21.655 | 20.972 | 20.252 | 19.110 | 21.812 |
| | AWMRmF | 24.109 | 23.543 | 23.069 | 22.431 | 21.819 | 21.149 | 20.448 | 19.298 | 21.983 |
| | AFMF | 28.966 | 28.348 | 27.670 | 26.667 | 25.939 | 24.840 | 20.785 | 13.921 | 24.642 |
| | AWMF | 30.201 | 29.437 | 28.741 | 27.850 | 26.979 | 26.037 | 24.660 | 22.619 | 27.065 |
| | ACmF | 30.314 | 29.499 | 28.778 | 27.872 | 26.986 | 26.042 | 24.662 | 22.622 | 27.097 |
| Paprika | ARmF | 30.556 | 29.652 | 28.926 | 27.981 | 27.061 | 26.074 | 24.679 | 22.621 | 27.194 |
| | IAWMF | 30.743 | 29.886 | 29.253 | 28.330 | 27.469 | 26.523 | 25.187 | 23.253 | 27.580 |
| | DAMRmF | 30.775 | 29.951 | 29.306 | 28.414 | 27.591 | 26.633 | 25.308 | 23.217 | 27.649 |
| | AWMRmF | 30.764 | 29.994 | 29.393 | 28.510 | 27.706 | 26.803 | 25.546 | 23.542 | 27.782 |

From Fig. 2, the PSNR graphs of the selected images are observed. Consistent with the previous findings, the AWMRmF filter yielded better results, further attesting to its superior noise removal capabilities.

In Fig. 3, there are the SSIM graphs. There was an improvement in AWMRmF filtering in comparison with others.

**Table 4** Mean SSIM results for some typical images with SPN ratios varying from 60 to 95%

| Images | Methods | 60% | 65% | 70% | 75% | 80% | 85% | 90% | 95% | Mean |
|---|---|---|---|---|---|---|---|---|---|---|
| | AFMF | 0.8944 | 0.8780 | 0.8607 | 0.8408 | 0.8136 | 0.7746 | 0.6889 | 0.3935 | 0.7681 |
| | AWMF | 0.9119 | 0.8986 | 0.8828 | 0.8641 | 0.8417 | 0.8114 | 0.7711 | 0.6973 | 0.8348 |
| | ACmF | 0.9132 | 0.8996 | 0.8836 | 0.8646 | 0.8421 | 0.8116 | 0.7711 | 0.6973 | 0.8354 |
| Lena | ARmF | 0.9162 | 0.9026 | 0.8864 | 0.8672 | 0.8443 | 0.8134 | 0.7725 | 0.6982 | 0.8376 |
| | IAWMF | 0.9183 | 0.9061 | 0.8917 | 0.8750 | 0.8539 | 0.8264 | 0.7897 | 0.7253 | 0.8483 |
| | DAMRmF | 0.9195 | 0.9073 | 0.8935 | 0.8773 | 0.8565 | 0.8294 | 0.7940 | 0.7260 | 0.8504 |
| | AWMRmF | 0.9195 | 0.9079 | 0.8944 | 0.8788 | 0.8585 | 0.8328 | 0.7997 | 0.7371 | 0.8536 |
| | AFMF | 0.9023 | 0.8883 | 0.8703 | 0.8501 | 0.8268 | 0.7942 | 0.6969 | 0.4123 | 0.7801 |
| | AWMF | 0.9233 | 0.9129 | 0.8972 | 0.8804 | 0.8603 | 0.8340 | 0.7961 | 0.7390 | 0.8554 |
| | ACmF | 0.9250 | 0.9143 | 0.8982 | 0.8810 | 0.8607 | 0.8343 | 0.7963 | 0.7390 | 0.8561 |
| Bird in the lake | ARmF | 0.9308 | 0.9198 | 0.9036 | 0.8861 | 0.8656 | 0.8384 | 0.7998 | 0.7415 | 0.8607 |
| | IAWMF | 0.9342 | 0.9241 | 0.9096 | 0.8936 | 0.8746 | 0.8496 | 0.8156 | 0.7642 | 0.8707 |
| | DAMRmF | 0.9326 | 0.9225 | 0.9083 | 0.8925 | 0.8746 | 0.8507 | 0.8172 | 0.7603 | 0.8698 |
| | AWMRmF | 0.9340 | 0.9245 | 0.9111 | 0.8960 | 0.8783 | 0.8552 | 0.8234 | 0.7725 | 0.8744 |
| | AFMF | 0.7987 | 0.7709 | 0.7439 | 0.7128 | 0.6756 | 0.6238 | 0.5410 | 0.3140 | 0.6476 |
| | AWMF | 0.8225 | 0.7976 | 0.7730 | 0.7424 | 0.7060 | 0.6606 | 0.6093 | 0.5313 | 0.7053 |
| | ACmF | 0.8245 | 0.7988 | 0.7742 | 0.7432 | 0.7065 | 0.6609 | 0.6095 | 0.5313 | 0.7061 |
| House | ARmF | 0.8339 | 0.8077 | 0.7827 | 0.7509 | 0.7129 | 0.6664 | 0.6136 | 0.5336 | 0.7127 |
| | IAWMF | 0.8394 | 0.8147 | 0.7914 | 0.7610 | 0.7254 | 0.6803 | 0.6297 | 0.5535 | 0.7244 |
| | DAMRmF | 0.8374 | 0.8130 | 0.7896 | 0.7598 | 0.7253 | 0.6812 | 0.6323 | 0.5541 | 0.7241 |
| | AWMRmF | 0.8413 | 0.8179 | 0.7955 | 0.7660 | 0.7321 | 0.6888 | 0.6407 | 0.5638 | 0.7308 |
| | AFMF | 0.8979 | 0.8826 | 0.8637 | 0.8389 | 0.8111 | 0.7703 | 0.6699 | 0.3757 | 0.7637 |
| | AWMF | 0.9174 | 0.9042 | 0.8883 | 0.8677 | 0.8450 | 0.8116 | 0.7657 | 0.6844 | 0.8356 |
| | ACmF | 0.9192 | 0.9057 | 0.8895 | 0.8684 | 0.8455 | 0.8119 | 0.7659 | 0.6845 | 0.8363 |
| Paprika | ARmF | 0.9235 | 0.9097 | 0.8937 | 0.8721 | 0.8488 | 0.8145 | 0.7678 | 0.6857 | 0.8395 |
| | IAWMF | 0.9261 | 0.9134 | 0.8994 | 0.8793 | 0.8585 | 0.8267 | 0.7848 | 0.7115 | 0.8500 |
| | DAMRmF | 0.9256 | 0.9132 | 0.8992 | 0.8794 | 0.8594 | 0.8284 | 0.7875 | 0.7122 | 0.8506 |
| | AWMRmF | 0.9252 | 0.9134 | 0.9003 | 0.8811 | 0.8621 | 0.8320 | 0.7936 | 0.7233 | 0.8539 |

**Table 5** Mean execution time for 26 typical images with SPN ratios varying from 60 to 95%

| Methods | 60% | 65% | 70% | 75% | 80% | 85% | 90% | 95% | Mean |
|---|---|---|---|---|---|---|---|---|---|
| AFMF | 10.215 | 9.807 | 9.418 | 9.196 | 9.220 | 9.448 | 10.187 | 11.891 | 9.923 |
| AWMF | 3.216 | 3.104 | 3.106 | 3.120 | 3.264 | 3.413 | 3.823 | 4.881 | 3.491 |
| ACmF | 1.075 | 1.139 | 1.249 | 1.340 | 1.519 | 1.645 | 1.833 | 2.245 | 1.506 |
| ARmF | 0.690 | 0.743 | 0.792 | 0.876 | 0.980 | 1.083 | 1.241 | 1.654 | 1.007 |
| IAWMF | 14.337 | 14.931 | 15.925 | 17.732 | 20.633 | 24.552 | 31.173 | 46.612 | 23.237 |
| DAMRmF | 4.268 | 4.895 | 5.684 | 6.668 | 8.017 | 9.855 | 12.645 | 18.011 | 8.755 |
| AWMRmF | 4.494 | 4.760 | 5.122 | 5.584 | 5.832 | 6.122 | 6.522 | 6.821 | 5.657 |

Figure 4 shows the image “Bird in the lake” original version, it’s noisy image and it’s filtered images with different filters.

In Fig. 5, the performance of AWMRmF filter with SPN density 75%, 85% and 95% is observed.

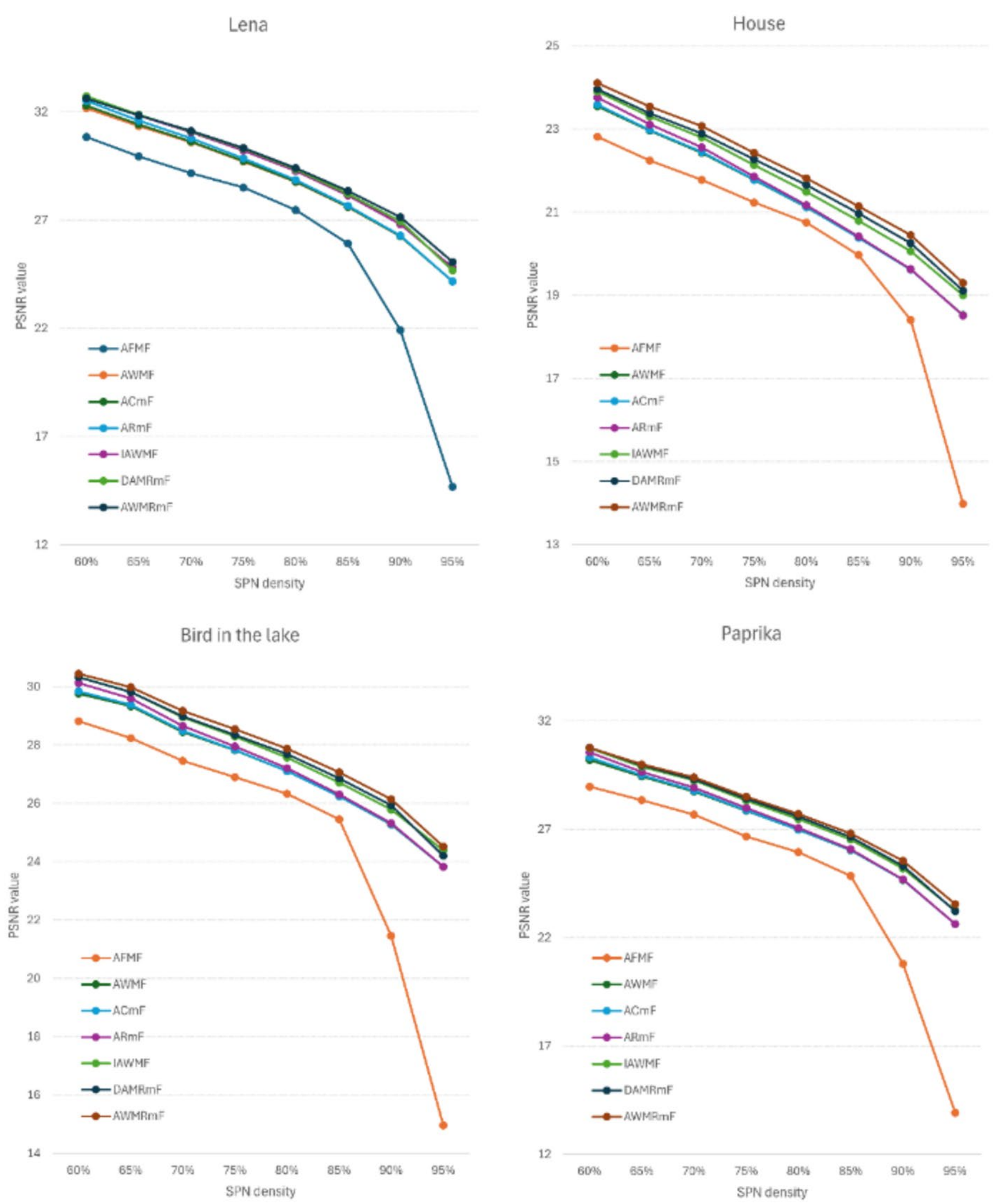

**Fig. 2** PSNR graphs for the selected images

## Comparison

Figure 6 represents the noisy images with SPN density 60%.The images"Heron" and"Waterlily", when processed with the DAMRmF filter, yielded significantly low PSNR values.

Figure 7 shows the result of filtering with the DAMRmF. This filter struggled to appropriately denoise the extremely dark regions marked in the images.

Upon applying the AWMRmF filter, there was a marked improvement in the PSNR values, as seen in Fig. 8. The AWMRmF filter handled the dark region better than the DAMRmF. This improved performance was mirrored in the SSIM graph, where the AWMRmF filter outperformed the DAMRmF filter, indicating its superiority in both metrics.

Figure 9 shows the PSNR and SSIM comparison graphs between DAMRmF and AWMRmF.

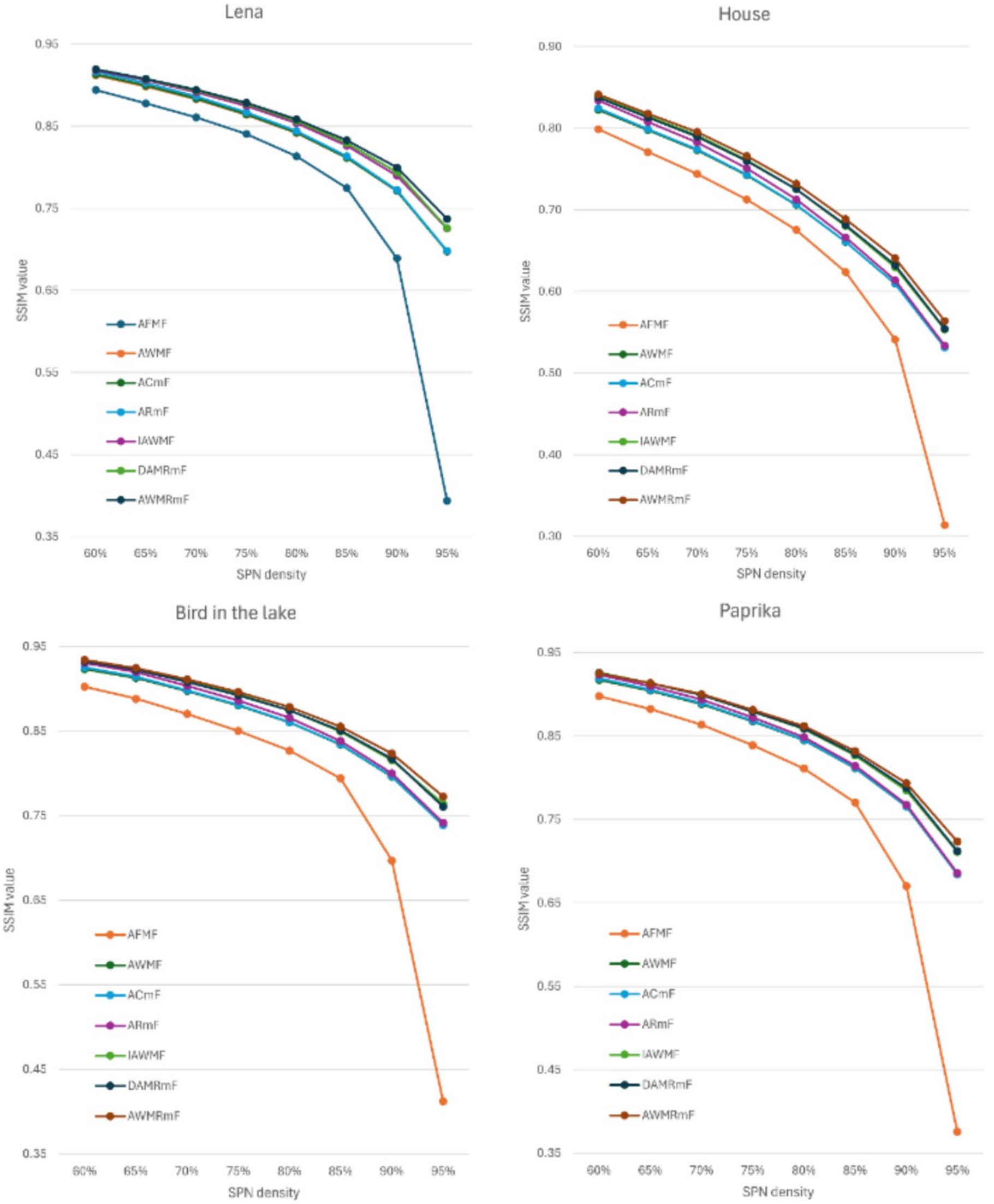

**Fig. 3** SSIM graphs for the selected images

**PSNR comparison with DAMRmF:**

Table 6 represents the PSNR values of two dark images which were filtered by DAMRmF and AWMRmF, respectively. For the dark images, AWMRmF performed better than DAMRmF. In DAMRmF, the PSNR values were very low in the extremely dark regions of the images. But the AWMRmF filter handled it easily and the PSNR values were very high.

**SSIM comparison with DAMRmF**

Table 7 represents the SSIM values of two dark images that have been filtered by DAMRmF and AWMRmF, respectively. Here, AWMRmF performed better than the DAMRmF.

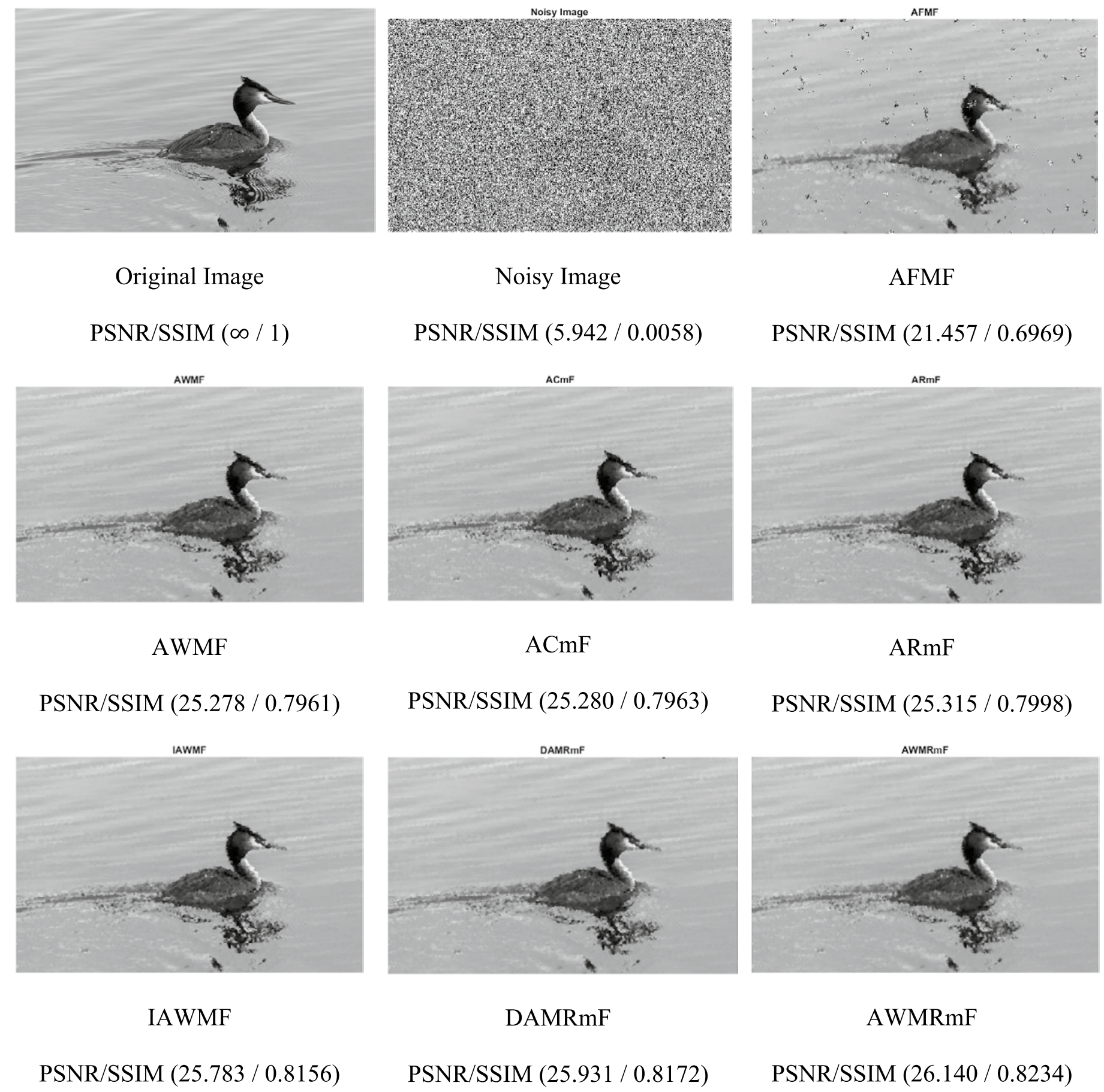

**Fig. 4** Filtered images with SPN level 90% of the image "Bird in the lake" (397 × 640 pixels)

## Conclusion

This study proposed a replacement of the weight of Riesz mean filter previous with a modified one and the function is weight-modified Riesz mean (WMRm). Moreover, I extended the iteration of the first loop. After applying WMRm function, the loop wasn't broken in the algorithm.

I designed a filter adaptive weight-modified Riesz mean filter (AWMRmF) for high-density SPN. Especially, when the SPN increased AWMRmF performed better than the other filters.

Moreover, AWMRmF ran faster than DAMRmF and IAWMF. Though AWMF, ARmF and ACmF ran faster than AWMRmF, those filters showed lower PSNR and SSIM values than AWMRmF for the selected images. Besides, AWMRmF performed better than DAMRmF in case of handling the extremely dark regions of the images.

Though AWMRmF has better performance than DAMRmF including other filters in denoising the SPN, it can be improved more by applying new pixel weight function and adaptivity conditions. Thus, we will be able to denoise highly-dense SPN from the image.

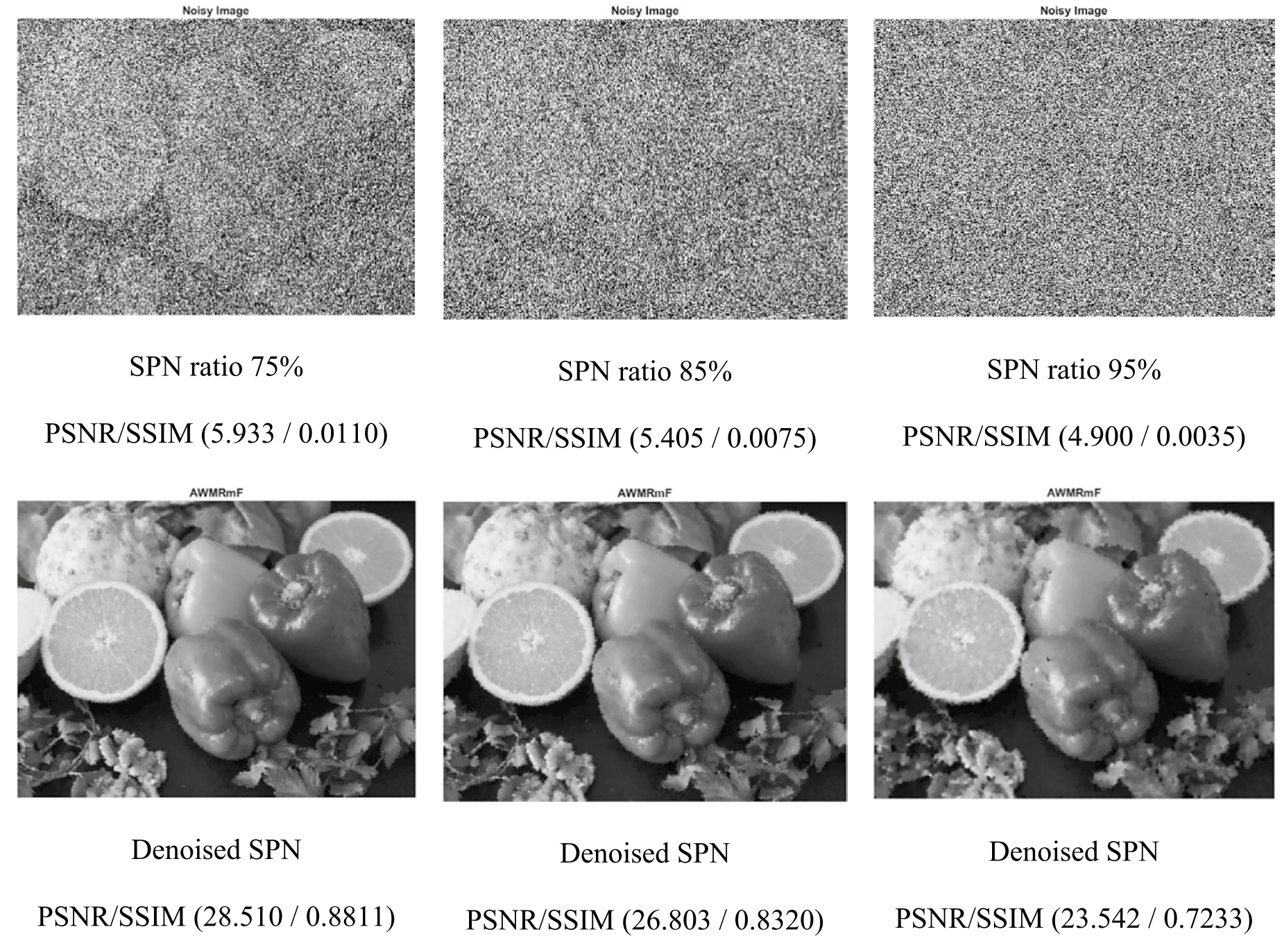

**Fig. 5** Images filtered by AWMRmF of the image "Paprika" (476 × 640 pixels)

A. *Comparison*

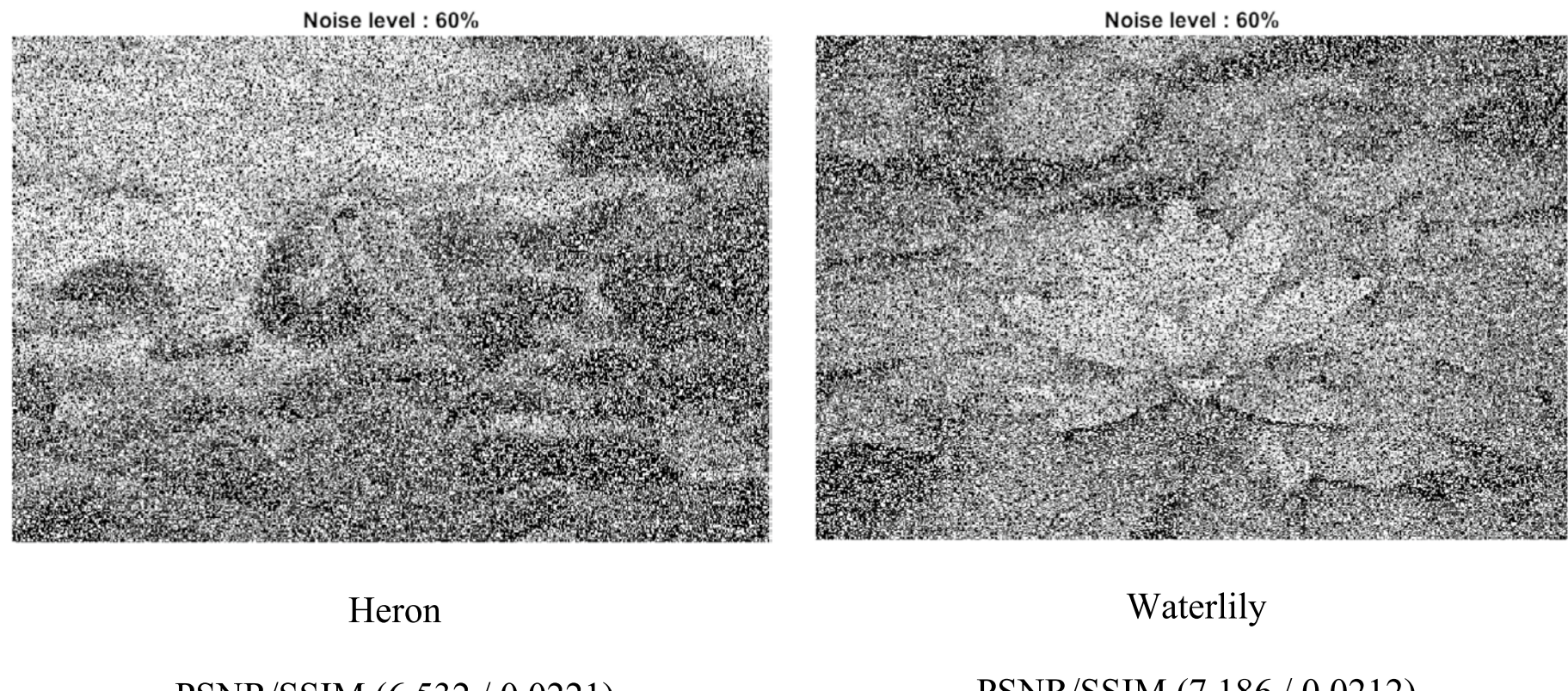

**Fig. 6** Noisy images with SPN density 60% (427 × 640 pixels)

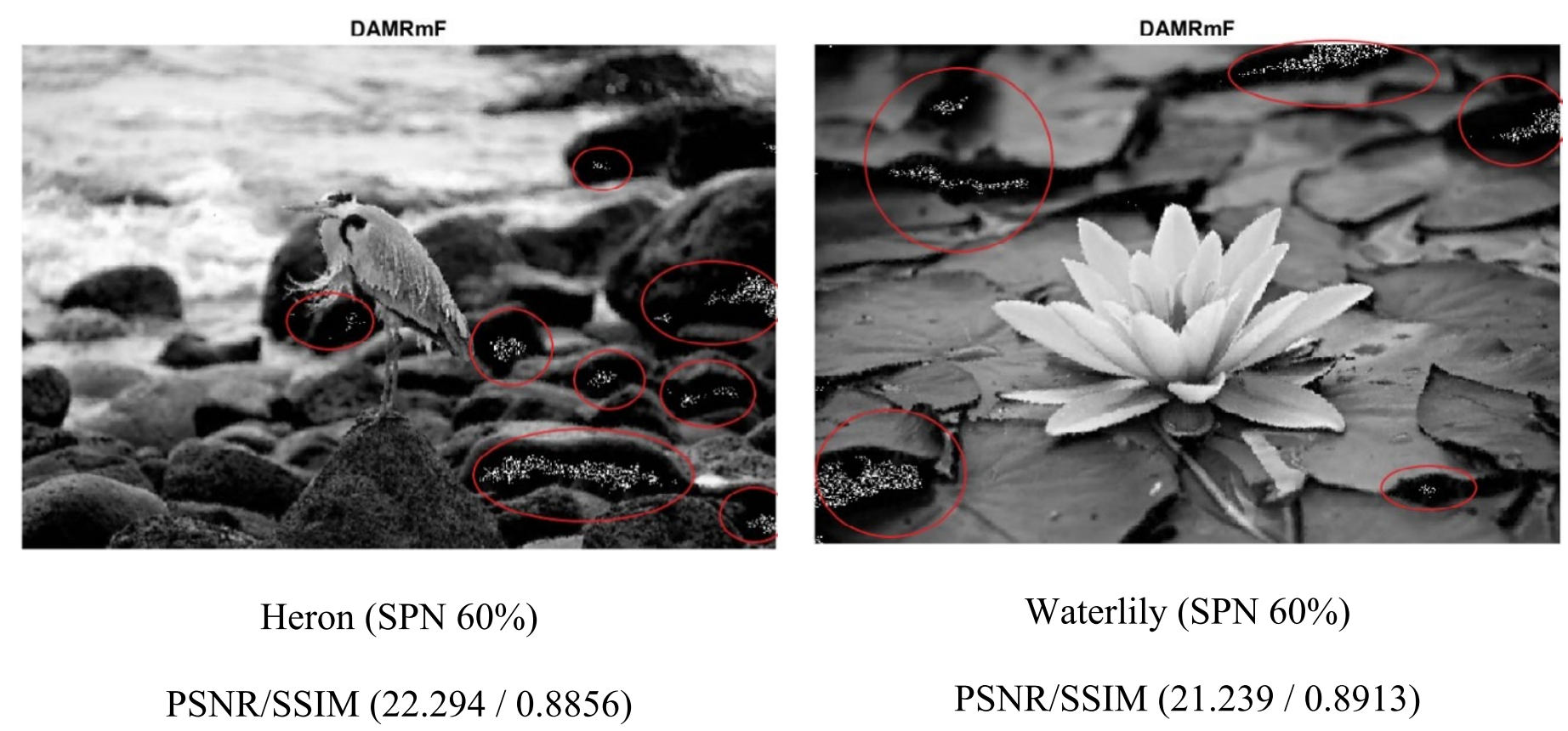

**Fig. 7** Filtered by DAMRmF of the selected images (427 × 640 pixels)

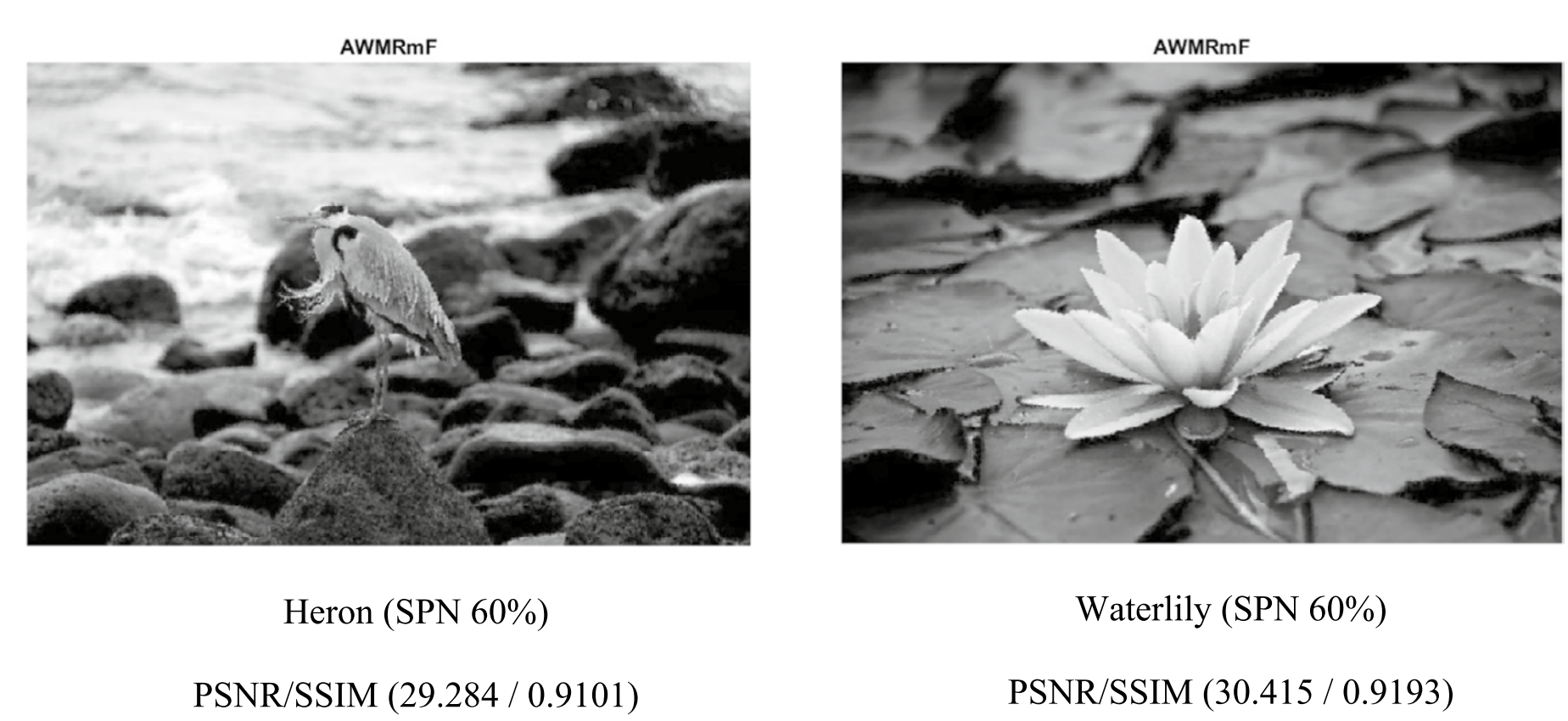

**Fig. 8** Filtered by AWMRmF of the selected images (427 × 640 pixels)

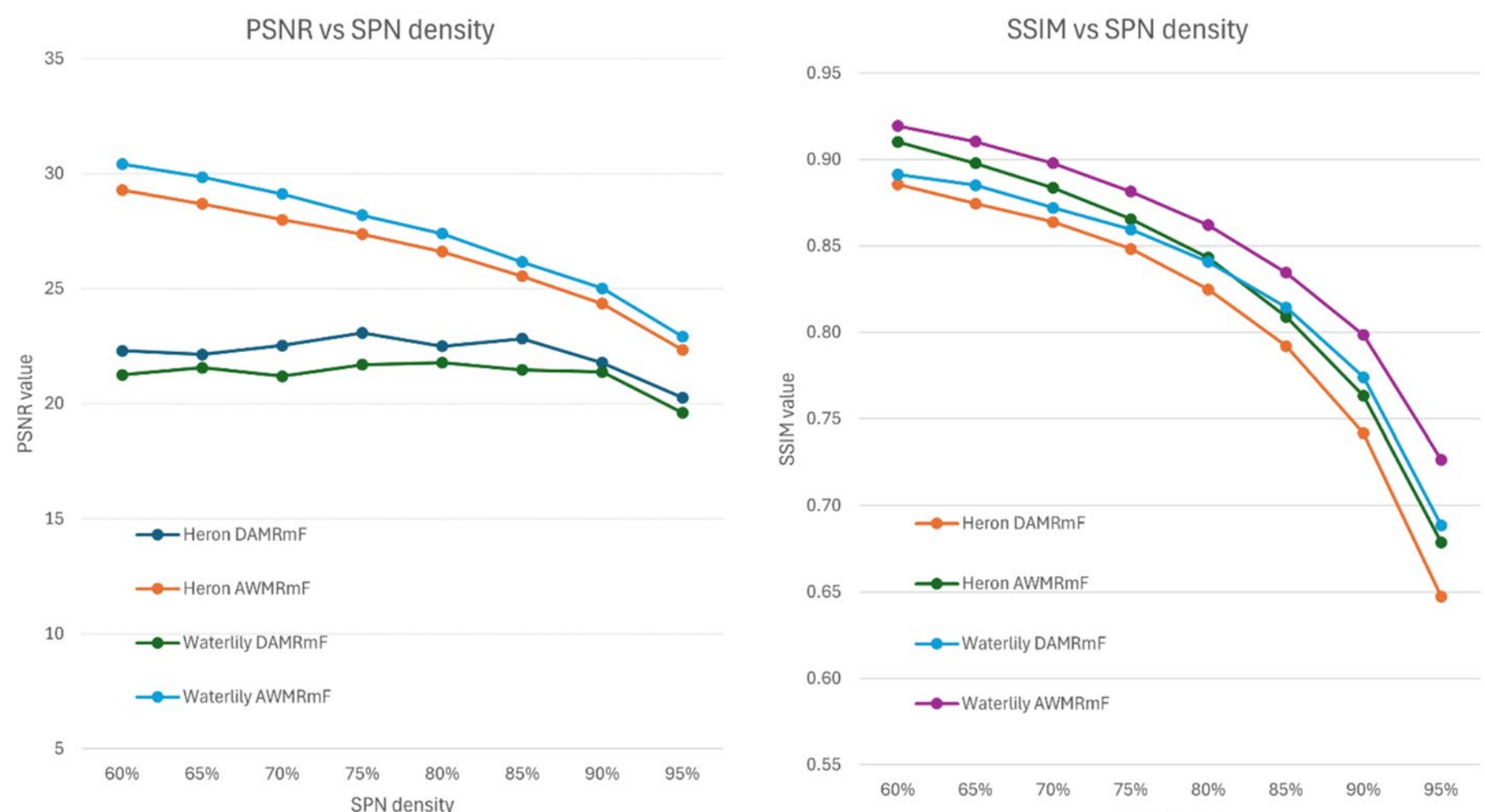

**Fig. 9** PSNR and SSIM graphs of the images "Heron" and "Waterlily"

**Table 6** PSNR values for extremely dark images with SPN ratios varying from 60 to 95%

| Images | Methods | 60% | 65% | 70% | 75% | 80% | 85% | 90% | 95% |
|---|---|---|---|---|---|---|---|---|---|
| Heron | DAMRmF | 22.294 | 22.129 | 22.520 | 23.071 | 22.483 | 22.814 | 21.761 | 20.253 |
| | AWMRmF | 29.284 | 28.678 | 27.989 | 27.363 | 26.596 | 25.533 | 24.340 | 22.328 |
| Waterlily | DAMRmF | 21.239 | 21.549 | 21.185 | 21.684 | 21.780 | 21.463 | 21.362 | 19.583 |
| | AWMRmF | 30.415 | 29.839 | 29.113 | 28.181 | 27.388 | 26.143 | 25.009 | 22.893 |

**Table 7** SSIM values for extremely dark images with SPN ratios varying from 60 to 95%

| Images | Methods | 60% | 65% | 70% | 75% | 80% | 85% | 90% | 95% |
|---|---|---|---|---|---|---|---|---|---|
| Heron | DAMRmF | 0.8856 | 0.8744 | 0.8638 | 0.8483 | 0.8247 | 0.7920 | 0.7417 | 0.6473 |
| | AWMRmF | 0.9101 | 0.8978 | 0.8836 | 0.8654 | 0.8433 | 0.8090 | 0.7633 | 0.6786 |
| Waterlily | DAMRmF | 0.8913 | 0.8851 | 0.8720 | 0.8594 | 0.8408 | 0.8144 | 0.7740 | 0.6884 |
| | AWMRmF | 0.9193 | 0.9104 | 0.8978 | 0.8815 | 0.8621 | 0.8345 | 0.7985 | 0.7265 |

**Abbreviations**

SPN Salt-and-pepper noise
IM Image matrix
NIM Noise image matrix

**Acknowledgements**
Not applicable.

**Author contributions**
Not applicable.

**Funding**
Not applicable.

**Availability of data and materials**
All the images used in the experiment taken from the site: https://pixabay.com/photos, The picture "Lena" was taken from the paper: https://dergipark.org.tr/en/pub/ejosat/issue/60692/873312.

## Declarations

**Competing interests**
The author declares that they have no competing interests.